\documentclass[%
 aip,
 amsmath,amssymb,
 reprint,%
]{revtex4-1}

\usepackage{graphicx}
\usepackage{dcolumn}
\usepackage{bm}
\usepackage{xcolor}
\usepackage[utf8]{inputenc}
\usepackage[T1]{fontenc}
\usepackage{mathptmx}
\usepackage{etoolbox}
\makeatletter
\def\@email#1#2{%
 \endgroup
 \patchcmd{\titleblock@produce}
  {\frontmatter@RRAPformat}
  {\frontmatter@RRAPformat{\produce@RRAP{*#1\href{mailto:#2}{#2}}}\frontmatter@RRAPformat}
  {}{}
}%
\makeatother
\begin{document}

\preprint{AIP/123-QED}

\title[MRI Driven Black Holes as Dark Matter]{MRI Driven Black Holes as Dark Matter:\\}

\author{B. Curd}
\affiliation{Department of Physics and Astronomy, The University of Texas at San Antonio, One UTSA Circle, San Antonio, TX 78429, USA \\}

\author{R. Anantua}%
\email{nate.lujan@my.utsa.edu}
\affiliation{Department of Physics and Astronomy, The University of Texas at San Antonio, One UTSA Circle, San Antonio, TX 78429, USA \\}
\affiliation{Physics \& Astronomy Department, Rice University, Houston, TX
77005-1827, USA}

\author{N. Lujan}
\email{richard.anantua@utsa.edu}
\affiliation{Department of Physics and Astronomy, The University of Texas at San Antonio, One UTSA Circle, San Antonio, TX 78429, USA \\}

\author{T.K. Fowler}%
\affiliation{Department of Nuclear Engineering, University of California, Berkeley, 4153 Etcheverry Hall, Berkeley, CA 94720 USA}

\date{\today}

\begin{abstract}
\textcolor{black}{}
We show scenarios in which primordial black hole accretion under the magnetorotational instability (MRI) uniquely relates the density of the early Universe to the abundance of present day dark matter.We demonstrate via long duration general relativistic magnetohydrodynamic (GRMHD) simulations that MRI-dominated accretion at least hundreds of gravitational radii from black holes can occur under conditions expected in the Positronium Era. We thereby identify that the positronium plasma that existed 0.01 s to 14 s into the Big Bang can serve as the primary source of mass that augmented primordial black hole seeds to $10^{16-17}$g black holes contributing to dark matter today. This population of black holes, in turn, radiates in a manner consistent with the observed gamma ray background. At a time of uncertainty about the role of new kinds of particles, the better understood primordial black hole MRI accretion process may be the best way to pin down how much dark matter mass lies behind horizons versus new dark sector particles. 
\end{abstract}

\maketitle

\section{\label{sec:level1}Introduction:\protect\\}
The dense, hot plasma of the primordial universe a few seconds after the Big Bang supported extreme rates of electron-positron pair production and sound speeds approaching the speed of light-- conditions that coupled with the presence of rotating primordial black holes may hold the key to trapping much of the present day dark matter mass behind their horizons. The assumption that dark matter comprises a new ``dark sector" of particle physics has not yet led to the discovery  of such particles \cite{Cooley2021,Slatyer2024}, while recent review papers by Carr and associates remind us that primordial black holes remain leading alternatives \cite{Carr2020,Carr2021}. Moreover, Carr and K\"uhnel conclude that the most likely mass range for these black holes is $10^{16 - 17}g$  \citep[][Sect. VII]{Carr2020}. In our work in \cite{Fowler2023}, we find that black holes as small as $10^{16 - 17}g$ were likely produced in the interval $0.01 - 14$ s after the Big Bang, when relativistic positronium (electrons and positrons) was the dominant contribution to primordial mass.

We arrive at this “Positronium Era” \textcolor{black}{(fully ionized or not)} as the time of interest as follows. Before $t = 0.01 \mathrm{ s}$ into the Big Bang, the temperature was $> 10$ MeV, not conducive to black hole formation. After $t = 14 \,{\rm s}$, all electrons and positrons had annihilated \cite{Sobrinho2024}, leaving as the only possibility for black hole formation the few protons and neutrons that did finally create stars and galaxies, but with less total mass than dark matter. Hence our focus on $0.01 \,{\rm s} < t < 14 \,{\rm s}$ when the positronium mass greatly exceeded what became dark matter. Accretion of only a tiny fraction of the positronium mass could account for all of dark matter. Moreover, only the formation of black holes could have preserved any of this annihilating positronium mass to the present era. 

To obtain an estimate of positronium dark matter, we need an accretion mechanism. For this, we turn to accretion by magneto-rotational instability (MRI), shown by  many authors to have contributed to accretion on galactic space scales \citep[][and references therein]{Colgate2014,Colgate2015}. As we shall see, MRI was likely the main source of black holes as dark matter and contributes $30\%$ or more of dark matter in our models.

The Positronium Era was right for MRI \cite{Fowler2023}. Ambient density fluctuations of order $\delta \rho = 10^{-5}\rho_{AMB}$ \cite{Tegmark1998} could confine growing seed masses. Seed growth was initiated by Bondi accretion \cite{Frank2002}, evolving into an MRI dynamo due to rotation in the primordial magnetic field \cite{latif2016, Balbus1998}.  Rotation could have occurred when adjacent irregular seed masses set each other in counter-rotation \cite{Fowler2023, Peebles1969}.  And average seed separations of order of the MRI accretion radius allowed uninhibited growth of the seed masses \cite{Fowler2023}.
    
While the importance of magnetic fields is well known \cite{latif2016}, we are not aware of comparable work on primordial MRI accretion and suggest that MRI accretion may be the most productive alternative to the search for new particles as dark matter. Given the present uncertainties about new particles, the better understood accretion process may be the best way to pin down how much and whether new particles must exist. If so, accretion by MRI will have contributed to two of five outstanding problems in astronomy and astrophysics identified in the 1999 Special Issue of Reviews of Modern Physics \cite{bederson1999}: ultra-high-energy cosmic rays \cite{Fowler2019}, and dark matter.

The paper is organized as follows. Our main result, for dark matter abundance, is presented \textcolor{black}{up front. Section II derives abundance using only fluid momentum and the primordial magnetic field in Ohm’s Law. The same result is derived} from the full MHD equations using the KORAL code, in Section III. Accretion masses are discussed in Section IV. \textcolor{black}{How accretion initiated by MRI far from the black hole extends to the black hole is discussed in Section V. Why accreting masses become black holes is explained in Section VI. Related work and a Summary are given in Sections VII and VIII.}

\section{\label{sec:level1}Dark Matter Abundance by Accretion:}
The goal of this paper is to calculate dark matter abundance due to accretion of primordial plasma, in the positronium era discussed in Section I.\textcolor{black}{We first show that Dark Matter abundance can be derived without assuming MRI transport, then identify MRI as the mechanism in Section III.}

\textcolor{black}{We define dark matter abundance as the ratio of accreted positronium mass $M \equiv M_{\rm DM}$ to the mass $M_0$ of protons and neutrons inside an accretion volume, radius $R_0$.} Given that there were $10^9$ electrons and positrons for each proton and neutron \cite{Fowler2023}, we obtain, with volume $V = (4\pi/3)R_0^3$ and masses $m_n \approx m_p = 1836 m_e$:

\begin{subequations}
\begin{align}
    \left( M_{DM}/M_0\right) &= \left[M/ (\rho V / 10^9m_e) (2 \times 1836 m_e)\right]\\
    &= 2.7 \times 10^5(M/(4\pi/3)\rho R_0^3) \\
    \rho t^2 &= 5\times10^5({\rm{g \, s^2 \, cm^{-3}}}) 
\end{align}
\end{subequations}
\textcolor{black}{where $\rho = ((7/4)/[1+ (7/4) + (21/8)]) \rho_{DE}$ \cite{Rafelski2023} is the positronium ambient density with radiation-dominated $\rho_{DE} = (3/8\pi {G})(10^5\rm{H/Mpc})$; Newtonian $G$; and CMB Hubble parameter $(H \approx 67.5  \ \rm km\  \rm s^{- 1} \rm Mpc^{-1})$ \cite{Plank2020}.}



\textcolor{black}{
It is believed that the primordial positronium medium supported dynamical magnetic fields \citep{latif2016}. We can estimate dark matter abundance given only the scaling $R_0 \propto (c_st)^{2/3}$ that follows from the conservation of magnetic helicity (twisted-ness) for magnetic waves excited by gravitational flow v toward a seed mass M [yielding magnetic flux $\Psi$ given by $\partial \Psi/\partial t +v \cdot \nabla \Psi \approx cRD(R)$ at R where hyper-resistivity $D \propto <\delta vx\delta B>$ passes through zero, [Ref.(8), Eqs. (48 - 49)].}

\textcolor{black}{
To derive dark matter abundance, the only information we need is the factor F in the following formulation of the helicity scaling law:}

\begin{align}
    R_0 = F(MG)^{1/3} t^{2/3}
\end{align}

\textcolor{black}{\noindent with $R_g = (MG/c_S^2 )$ and thermal velocity $c_S$ . Introducing this $R_0$ into Equation (1b) gives:}

\begin{align}
    (M_{DM}/M_0) = 2.7 \times 10^5 (M/[4\pi/3]\rho R_0^3) = 2\times 10^6 (1/F^3)
\end{align}

\textcolor{black}{\noindent where $M$ in the numerator cancels $M$ in $R_0^3$. A value $F = 70$ gives $(M_{DM}/M_0)=6$, $100\%$ of dark matter. A value $F = 100$ suggested by extrapolation of Figure 1 to higher R and t gives $(M_{DM}/M_0)= 2,$  or $30\%$ of dark matter. We will adopt $F=100$ in this paper.}


\textcolor{black}{Only constant $M$ contributes to gravity in KORAL. Integrating mass flow to determine $M$ would give a different estimate of dark matter abundance, using the standard mass accretion equation in Frank, King and Raine (Chapter 2)\cite{Frank2002}:}

\begin{align}
    dM/dt = - 4\pi \rho R^2 v
\end{align}
\textcolor{black}{\noindent Here magnitudes $v = c_S\approx c $ and $R \approx  2R_g = R_S$ (Schwarzschild) gives Bondi accretion (Chapter 2)\cite{Frank2002}, while $v = A c_S (R_g/R_0)^{1/2}$ and $R = R_0$ gives MRI with $0.001 < A < 0.01$ fitting $v$ to the shaded region in Figure 1, derived in Section III. The main difference is that the large MRI accretion range $R_0 \gg R_S$ more than compensates for the slower accretion velocity by MRI, so that MRI accretion dominates over Bondi accretion when MRI occurs. } 

\textcolor{black}{Applying $\int ^tdt’$ to Equation (4) leaving $M(t)$ in the integrand unspecified gives: }


\begin{subequations}
\begin{align}
    {M_{\text{DM}}}/{M_0} &= 2.7 \times 10^5 \left[ \int^t dt'4 \pi \rho A(MG)^{1/2}R_0^{3/2}/ (4/3) \pi \rho R_0^3 \right] \\
    &= [8\times 10^5(A/F^{3/2})\int^t(dt'/t)(t/t')(M(t')/M)]
\end{align}
\end{subequations}
\textcolor{black}{\noindent where we introduce $\rho t^2$ from Equation (1c).}

\textcolor{black}{We expect Equations (3) and (5b) to agree. To evaluate Equation (5b), we integrate Equation (4) with $v_{MRI} = A(MG/R_0)^{1/2}$ and $R = R_0$ in Equation (2). After some algebra, using also $\rho$ from Equation (1c), we obtain $d/dt (\ln M) = C^*/t$ integrating to $M(t')\propto(t'/t)^{C^*}$ with $C^* = 0.4(AF^{3/2})$. With this $M(t’)$, Equation (5b) can be integrated, giving $(M_{DM}/M_0) = (2\times10^6/F^3)[C^*\int^t(dt'/t)(t'/t)^{c^*-1}] = 2\times 10^6/F^3$ in agreement with Equation (3). We adopt Equation (3) as the simpler result, yielding $(M_{DM}/M_0) = 2$ for $F=100$, independent of M and $C^*$. As noted earlier, $(M_{DM}/M_0) = 2$ accounts for $30\%$ of dark matter. Further refinements would require  longer computer runs to better pin down the parameter F.}


\section{\label{sec:level1}GRMHD Simulations}

\textcolor{black}{
In Section 2, we derived dark matter abundance from magnetic helicity without needing to deal with a complete fluid theory.  Here we ask whether a plasma fluid code would give similar results plus useful details. For this, we turn to the KORAL MHD code already applied to accretion in other contexts \cite{Sadowski2014, Sadowski+2015}. Simulating accretion all the way from $R_0$ to the black hole challenges the state-of-the-art KORAL code, limiting the scale to $R_0 > R > 100a$ for black hole radius $a \approx(MG/c^2)$. Even so, accumulating enough running time limits most of our simulations to 2D, shown to be equivalent to 3D tests in our domain.}

\textcolor{black}{
Results are given in Figures 1 – 7. Table 1 below (derived from Figure 1) yields scaling like Equation (2), hence supporting our model. That this scaling concerns magneto-rotational instability (MRI) is suggested by the conditions for MRI in Figure 2 and effects of magnetic fluctuations in Figures 3, 4, 5 \cite{Balbus1998, Fowler2023}. That instability ejects angular momentum allowing accretion to proceed is seen in Figure 6, though in space a Kepleriann disk could be sustained by closure of jet paths recycling angular momentum through the disk.}

\begin{table}[h]
    \renewcommand{\arraystretch}{1.4} 
    \setlength{\tabcolsep}{8pt} 
    \centering
    \begin{tabular}{|c|c|c|}
        \hline
        $t/t_g$ & $R_0/R_g$ & $2.5 (t/t_g)^{2/3}$ \\ \hline
        $10^5$ & $5 \times 10^3$ & $5 \times 10^3$ \\ \hline
        $10^6$ & $10^4$ & $2.5 \times 10^4$ \\ \hline
        $10^7$ & $10^5$ & $1.2 \times 10^5$ \\ \hline
        $5 \times 10^7$ & $10^6$ & $3.4 \times 10^5$ \\ \hline
    \end{tabular}
    \caption{Table verifying that r vs. t in the shaded zone of Figure 1 scales as $r/r_g  \propto t^{2/3}$ as argued in the text, giving r $\equiv$ $R_0$ in Equation (2) as the
    accretion radius growing with time. Figure 1 plots v vs. r at color-coded times so that r vs. t treats the sequence of colors as a time dimension. 
}
    \label{tab:placeholder_label}
\end{table}

\textcolor{black}{ Figure 2 shows the region in R – M space where MRI could have been initiated, scaled to $R_g$. The dimensions of this MRI zone were small compared to $R_o$; large compared to $R_g$; and order of collision mean free paths. Coulomb collisions dominated, creation and annihilation exchanging mass momentum and radiative momentum having little effect on accretion flow until annihilation dominates at 
$t = 14 \rm s$ [5]. Collisions, with mean free paths $\approx 1000 R_g$  (at t = 1 s, T = 1 MeV), smoothed-out the primordial pressure, while MRI turbulent transport helped preserve this naturally-flat ambient pressure profile. Constant pressure yielded the simplest MRI result in Balbus and Hawley ([12], Equation (111)). The smooth-average Debye length does not appear; MRI only requires $k_Rv_A = (v_A/R) < \Omega$ (with Alfven velocity $ v_A$, rotation $\Omega$). Given this, there are three other conditions for MRI accretion to occur:}
\textcolor{black}{
\begin{enumerate}
\item First, accretion must cause the ambient magnetic field to grow, from a seed magnetic field of order $\delta B \approx 1$ gauss to achieve the required MRI field $B(t) = \delta B \exp(\omega_{\rm MRI} t)$ in the time available. One gauss seed fields can be inferred from $10^{-15}$ gauss today \cite{latif2016};
\item MRI must grow faster than the Universe expansion scale ($\Omega t > 1$) (Curve I);
\item viscous flow $k(c_s^2/\gamma_C)$ must be slower than the MRI accretion velocity $v_{\rm MRI} \approx R\Omega$ (Curve II). The exact boundaries vary with the collision frequency $\gamma_C$ (see Collisionality, end of this Section).
\end{enumerate}}

\textcolor{black}{A scenario accounting for accretion from the MRI zone to the black hole is discussed in Reference \citep{Fowler2023}, Section 4. At first, constant $v_RR^2$ preserves $\nabla \cdot nv = 0$ giving constant n and T at their primordial values. When finally $v_R \propto R^{-2} \rightarrow c$, spherical convergence forces n to grow in competition with electron-positron annihilation. This forces $kR >> 1$ representing collapse into shells of mass too thin to annihilate.  
The other important feature is propagation of MRI-driven electromagnetic waves outward toward $R_0$. In this, the MRI zone acts as a radiating antenna. Here we deal with conditions allowing MRI-driven radiation to propagate.}

\textcolor{black}{
How MRI radiation was sustained is described by the generalized dispersion relation in Reference 5 including coupling of MRI to Alfven waves. Then combining mass flow v and momentum equations in cylinder (R, z) geometry gives, for particle density n, Bondi radius $\approx R_g$, frequency $\omega$, rotation $\Omega$, gravity (MG/R), Alfven velocity $v_A$, net positronium annihilation rate $\gamma_A$. and collision frequency $\gamma_C$:}

\begin{align}
\frac{\partial^2 n}{\partial t^2}
&\approx \omega^2 n
\approx -n\,\frac{\partial^2}{\partial R^2}
\left[\frac{M G}{R} + v^2\right]
- \frac{\partial (n \gamma_A)}{\partial t}\\
\left(\omega + \tfrac{i}{2}\gamma_A\right)^2 &\simeq (k_z^2 v_A^2 - 3\Omega^2) - \tfrac{1}{2} k_R^3 c^2 R_g C_1 + \left(\tfrac{i\gamma_A}{2}\right)^2\\
\frac{\partial \mathbf{B}}{\partial t} &= -c\,\nabla \times \mathbf{E} = \nabla \times (\delta \mathbf{v} \times \delta \mathbf{B})
\end{align}

\textcolor{black}{
for $C_1 = 1 + (\gamma_Ck_{R}v/ 1/2 {k_{R}}^3c^2R_g)$ and velocity and magnetic perturbations  $\delta v$ and $\delta B$. Taking n $\propto exp (i\omega t)$ and adding MRI terms gives Equation (7) \cite{Balbus1998}. We neglect pressure, since MRI can propagate at constant pressure. Also, we do not need to deal with radiation, since momentum exchange between radiation and accreting particles tends to preserve accretion flow v.}

\textcolor{black}{
MRI driven by rotation comes from the first term on the right in Equation (7). MRI generated near the black hole (Figure 2) propagates radially to $R = R_0$ that serves as an O-point around which poloidal flux $(B_R, B_z)$ circulates, analogous to kink-mode current drive observed in laboratory spheromaks [Ref. [7], Appendix B]. To give enough time for accretion, we focus on Big Bang time $1s < t < 14 s$ (significantly relativistic, $1\ \rm Mev > T > 300\ \rm keV$, when $\gamma_C \approx \gamma_A \approx  (3/8)(n\sigma_ Tc)$ (Thomson $\sigma_T$, \cite{Krolik1999})). Propagation of MRI works if Alfven wave lengths $k_R^{-1}$ satisfying Equation (7) fit within boundaries $v_\omega t = (\omega t/k_R)  > R$, giving $k_RR_g > [\gamma_ARR_g/(c^2t)C_1(1 + \rho_{AMB} (\gamma_C c/k_R^2R_g)^{1/2})]^{1/2} > 1$ with $C_1 = 1 + (\gamma_CR^2/cR_g)$ [Reference \cite{Fowler2023}, Equation (3c)]. After some algebra, using also $\rho_{AMB}$ in Equation (1c), we find that MRI in Figure 2 should have propagated to $R_0$ in the period of interest.}

\textcolor{black}{
Collisions also determine the applicability of the KORAL Magneto-Hydrodynamic (MHD) code to the primordial domain. KORAL is a relativistic MHD code, derivable from Boltzmann phase space distributions coupled to electromagnetic fields (the Vlasov equation). The main error in approximating the Vlasov equation by MHD arises from averaging over assumed Maxwellian momentum distributions f(x,p,t) to determine fluid parameters such as particle density n and temperature T (Ref. [20] Chap. 1).  Helicity conservation giving our result for dark matter abundance is independent of n and T, depending instead on gravitational acceleration by the seed mass M.}
\textcolor{black}{
Collisions among particles and radiation fields can favor quasi-Maxwellians.}

\subsection{Collisionality}

\textcolor{black}{Collisionless plasmas are frequently modeled by GRMHD simulations. For example, the M87 jet/accretion flow/black hole system has an electron-positron mean free path of $\lambda_\mathrm{mfp}\approx(k_BT_e)^2/(\pi n_e e^4)\approx 10^{19}T_{e,10}^2n_{e,6}^{-1}\approx(10^4r_g)$. \cite{Luepker2025} and has been well modeled to compare with multiwavelength data up to Event Horizon Telescope 230 GHz scales (e.g. Refs. [24, 25]). In such settings, turbulent heating can be governed by long range phenomena such as Alfv\'en waves, and short range forces neglected as Coulomb collision timescales far exceeding dynamical time scales for accretion.}

\textcolor{black}{Particle-in-cell (PIC) simulations have long reproduced collisionless MRI \cite{Riquelme2012} and has even been applied to MRI analysis in pair plasma \cite{Sandoval2024,Bacchini2024}. In analytical models\cite{Rosin2012}, linear instability analysis in the (weakly) collisional Braginskii MHD regime with the input of vertical fields into a plasma with a galactic rotation profile have recovered the MRI (albeit with a reduced growth rate). 
}

\textcolor{black}{
A non-relativistic estimate of the collision time in primordial electron-positron plasma is $\tau_\mathrm{coll}\simeq 1/(n_\gamma\sigma_Tc)\simeq9.1\times10^{-20}t^2\mathrm{s}$. However, this is an 
underestimate due to the 
relativistic nature of our plasma, in which the Thompson cross section gives way to the (far less efficient \cite{Heinzl2022}) Klein-Nishina cross section for photon-lepton scattering, and Moeller cross section \cite{Moeller1932} for electron-electron scattering-- which goes like $E^{-2}$.  }

\textcolor{black}{For example, define the relativistic limit $\gamma \beta\gtrsim 1$, i.e., $\beta=\frac{v}{c}\gtrsim 0.6$. Then most energetic non-relativistic electron energy is $0.25$ MeV. The MeV electrons in our simulated positronium plasma (moving $\sim .94c$) then have lepton scattering suppression by a factor of at least 16. }

\subsection{The KORAL Code}

\textcolor{black}{We simulate MRI using the KORAL code, the MHD code used to model MRI in other contexts. The KORAL code is fully described in \cite{Sadowski2013,Sadowski2014}. The code is also capable of evolving the radiation field \cite{Sadowski2013,Sadowski2014} and implementing a dynamo for 2D accretion disk problem \cite{Sadowski+2015}. Phenomena of interest are covered including flux compression in \cite{Igumenshchev2002, McKinney2012}. }

\textcolor{black}{Here we use KORAL to calculate accretion as the MRI accretion radius $R_0$ grows in time, starting near the black hole. We evolve the fluid in two dimensions $(r, \theta)$ in modified Kerr-Schild coordinates. The radial grid cells are spaced logarithmically, and we choose inner and outer radial bounds $R_{\rm min}=10^2 GM/c^2$ and $R_{\rm max}=10^6 GM/c^2$. Since the inner radial boundary is at $100 \, GM/c^2$, the effects of general relativity are negligible and we also cannot comment on the fate of gas beyond this point. However, this trade off was necessary to achieve the extremely long run times needed to study MRI well beyond the Bondi radius in the problem. We use nearly a full $\pi$ in polar angle $\theta$ and only cut out a region of $d\theta=0.005\pi$ at the top and bottom poles to ensure an efficient time step. We choose outflow boundary conditions at both the inner and outer radial bounds and reflective boundary conditions at the top and bottom polar boundaries. In each simulation, we employ a resolution $N_r\times N_\theta= 320\times 160$. Only the gravity of a Schwarzschild black hole of fixed mass $M$ is included.}

\begin{figure}
    \centering
    \includegraphics[width=1\columnwidth]{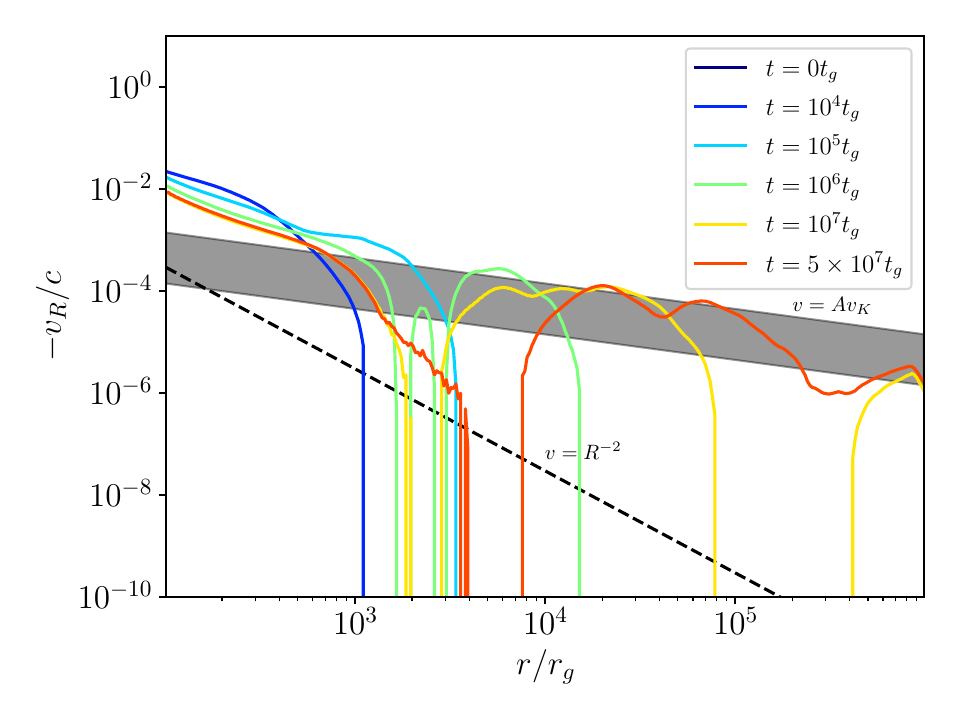}
    \caption{\label{fig1}
Showing the accretion velocity $-v_R$ versus spherical radius $r$ at various times (units: $\mathrm{ r_g=GM/c^2;\, t_g=r_g/c}$). Note peaks due to MRI. Actual $v_R$ is negative (see text). The shaded region shows $v=Av_K$ for values $0.001 < A < 0.01$.}
    \label{fig:placeholder}
\end{figure}

\textcolor{black}{We initialize the simulation with a fluid of constant gas density, constant gas temperature, equation of state $\gamma=4/3$ for a radiation dominated plasma, and Keplerian rotational speed, $v_{\phi}=v_K$, to excite the MRI. The choice of initial gas density $\rho$ is arbitrary since GRMHD is scale free. We set $v_r=v_{\theta}=0$. To evolve a fluid with similar properties to the positronium plasma in the early universe, we evolve a pure hydrogen gas but choose the temperature such that the sound speed is the same as that for  a positronium plasma. The gas temperature we set is 1836 times the temperature of an analogous positronium plasma with $T=10^3$ keV. The expected solution can be shown to be similar in the Bondi case, ignoring any effects from annihilation or magnetism, since the sound speed is the determining factor.}

\begin{figure}[ht]
\centering
\includegraphics[width=.95\columnwidth]{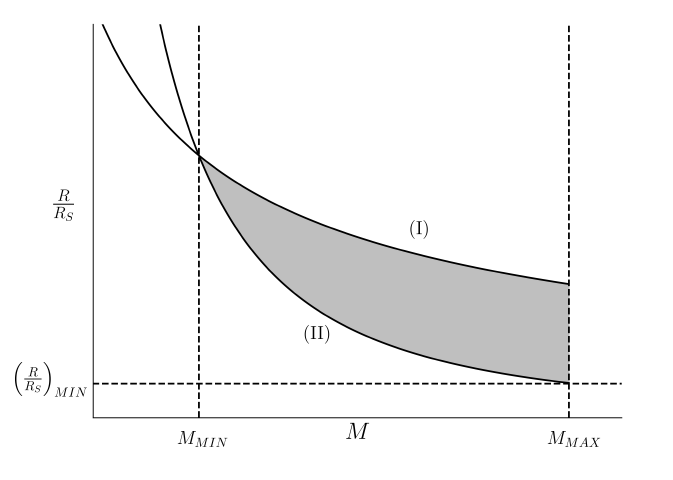}
\caption{\label{fig2}
Showing approximate boundaries in $R-M$ space where MRI could have been initiated at Big Bang  time $= 1\mathrm{\ s}$. The allowed space is shaded, with limits $M_{\mathrm{MIN}} = 10^{15} \mathrm{\ g}$; $M_{MAX} = 10^{18} \mathrm{\ g}$: and upper boundary (I) $\Omega t > I$ required for MRI to grow from noise, and lower boundary (II) for MRI flow to exceed the viscous flow rate (see Text).
}
\end{figure}

\textcolor{black}{We take the seed magnetic field generating MRI to be poloidal (cylindrical $B_r,B_z$) and initialize the field such that the gas to magnetic pressure ratio $\beta=p_{\rm gas}/p_{\rm mag}\approx 10^3$ at the innermost radial boundary. While pure $B_z$ creates MRI in the analysis in Ref. [12] (Sect. IV.B),  we found that an initial $B_r/B_z  \approx$ 0.1 expedites MRI development in KORAL. Due to the anti-dynamo theorem \cite{Cowling1933}, 2D accretion disk problems typically make use of a dynamo to sustain long term accretion \citep[e.g.][]{Sadowski+2015}; however, an artificial dynamo is not necessary to maintain accretion as we found the poloidal magnetic field is dragged in with gas from larger radii and the field does not decay over the run times we consider. As we discuss later, this is verified in 3D. The parameter range is challenging. Applying AGN formulas in Refs. [7,8,15] gives (for typical M = $10^{16}$ g, $t$ = 1s and $T$ = $10^3$ keV) primordial jets of length $\gg 10^8$ cm, radius $a = 1$ cm (resistive) and magnetic field $10^{10} (a/R)$ gauss. In “natural” units ($r_g = MG/c_s^2$ and $t_g = r_g/c_s$ with sound speed $c_s \approx c$), one second of accretion time is $10^{23}$ $t_g$. Our simulation runs up to $t = 5 \times 10^7$ $t_g$ are limited to the formation of an MRI jet current near the black hole where most of the accretion power is deposited \citep{Colgate2015}. In order to adequately resolve MRI in KORAL, we ensure that the fastest growing mode of the MRI is resolved via the MRI quality factor \citep{Hawley2011} $Q_\theta = \frac{2\pi}{\Omega\, dx^\theta}\frac{|b^\theta|}{\sqrt{4\pi\rho}}$, where $dx^\theta$ is the grid cell length in $\theta$ and $b^\theta$ is the magnetic field component in $\theta$. $Q_\theta$ must exceed 5 at a minimum to resolve the MRI. Our choice of initial $\beta$ is such that $Q_\theta \gtrsim 100$ for the duration of the simulation.}

\begin{figure}[htbp]
\includegraphics[width=\columnwidth]{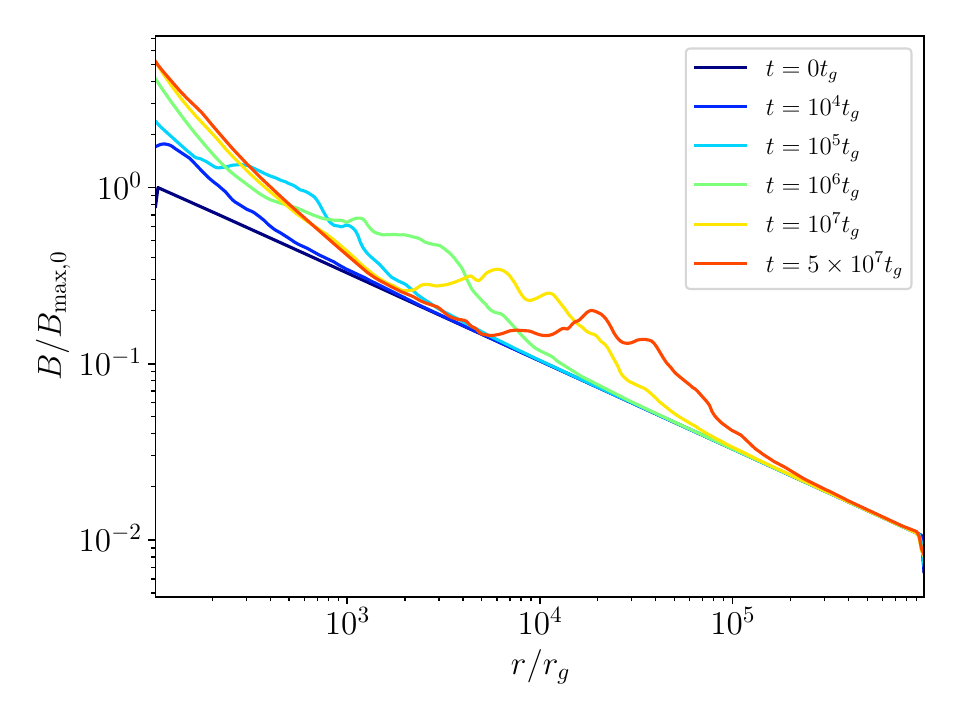} 
\caption{\label{fig3}
Showing the magnetic field $B$ versus $R$ at various times. Note 
correlation of oscillations in $B$ with oscillations in $v_R$ in Figure 1. Curves are scaled by the maximum field strength at the initial state, $B_{max,0}$.
}
\end{figure}

\textcolor{black}{The time evolution of the simulation demonstrates that growth of the accretion radius is correlated with oscillations in the magnetic field (Figures 3 and 4). As is typical in MRI driven accretion, the plasma becomes turbulent, as evident by the velocity streamlines (Figure 5). Angular momentum transport leads to an outflow at the mid-plane in Figure 5, and this was also found in hydrodynamics simulations using a parameterized viscosity $\alpha$ \citep{Li2013}. However, the net flow is towards the BH (Figure 1).}

\begin{figure}[htbp]
\includegraphics[width=0.95\columnwidth]{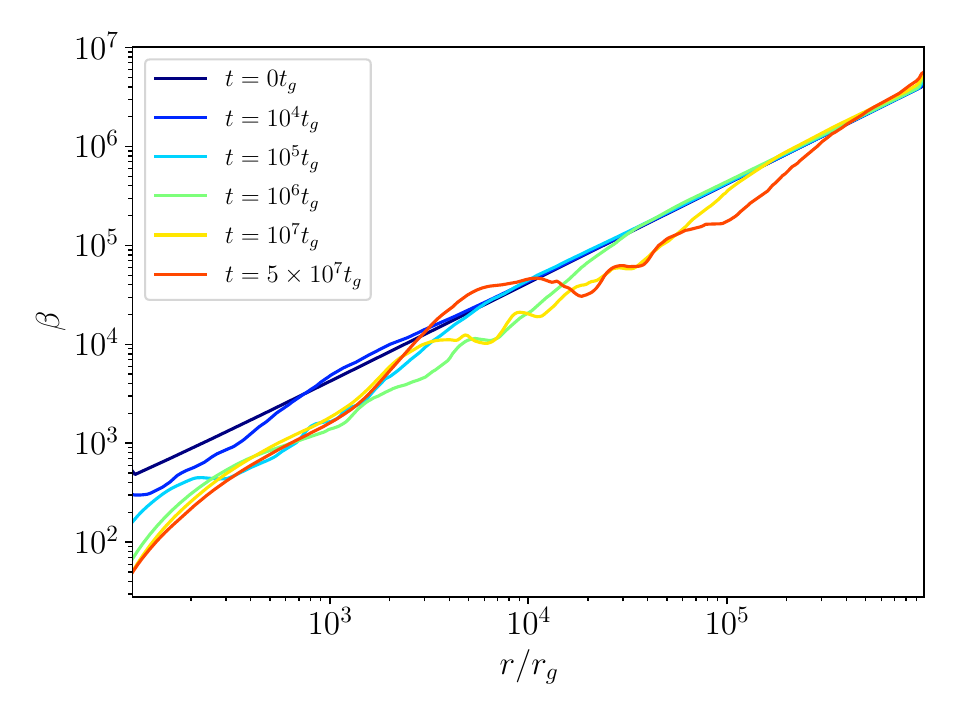}
\caption{\label{fig4} 
Showing $\beta$  the ratio of plasma pressure to magnetic pressure.
}
\end{figure}

\textcolor{black}{Accretion at $R_0$ is not much affected by accretion before and after $R_0$. In Figure \ref{fig1},
accretion at $R < R_0$ is due to ejection of angular momentum by the MRI magnetic jet cylindrical ($\partial(\rho r^2\Omega)/\partial t = (1/4\pi)\partial(rB_{\Phi}B_z)/\partial z + ...$) (Eq. (5))\cite{Colgate2014}. This leaves weakly-rotating plasma behind the MRI accretion wave, as seen in Figure \ref{fig6}. Weak rotation kills MRI, so that in KORAL MRI is localized around the advancing $R_0(t)$, while in general $v_R$ at $R_0$ may be due to propagation from MRI in the interior \citep{Fowler2023}. Weak rotation inside $R<R_0$ gives Bondi accretion in a magnetic field discussed in \cite{Cunningham2012}. This scenario is unlikely in space where rotation can be sustained by recycling angular momentum via large MRI jet current loops closing on themselves \citep{Colgate2014,Colgate2015}. The more likely scenario in space would be Jeans gravitational instability.}

\textcolor{black}{We verify that our primary result, that MRI increases the accretion radius (Figure 1), is not dependent on our choice of magnetic field strength by rerunning the problem with $\beta\approx 10^4$ and $10^5$ at the innermost radial boundary. The evolution and scalings are qualitatively the same, but with slightly slower propagation of the accretion front. We performed a short duration test ($t_{\rm max}=10^6 t_g$) of the $\beta\approx10^3$ model in 3D on a $N_r\times N_\theta \times N_\phi= 320\times 160\times 16$ grid, using a $\pi/2$ wedge in $\phi$ to reduce the resource requirement. We find excellent qualitative agreement for all features shown in Figure 1 and Figures 3-6, verifying that 2D is sufficient. We also verify that the Bondi accretion problem with the same plasma properties, but no magnetic field or rotation, in KORAL returns the expected $v_R\propto R^{-2}$ scaling.}
\newline

\section{\label{sec:level1} Dark Matter Masses}
Dark matter masses are obtained from Figure 2. The upper curve guarantees $v < R \Omega$ (MRI growth rate $<$ Keplerian $\Omega$)\cite{Balbus1998}. The lower curve is the transition from viscous flow to MRI at $R_0 (k^{2}c_S^2/\gamma_c\Omega) = 1$ at $kR = 1$ with collision frequency $\gamma_C \approx (10^{10} m_e /ln \Lambda\rho )T^{3/2}$ where ln $\Lambda = 20$, $T = (10^3/t^{1/2} keV$ (cgs/keV units)) \cite{Fowler2023}. Hawking mass evaporation limits accreted masses surviving today to $M > 10^{15} g$ \cite{Carr2020}, while the transition from viscous flow to MRI limits $M < 10^{18} g$ \cite{Fowler2023}. These limits are plotted in Figure 2, giving:

\begin{align}
    10^{15}g < M< 10^{18}g
\end{align}

This range is similar to the lowest “mass window” in (Fig. 1)\cite{Carr2020}, verified as being due to Hawking radiation in Ref.[5], Section 4.

\begin{figure}[htbp]
\includegraphics[width=0.95\columnwidth]{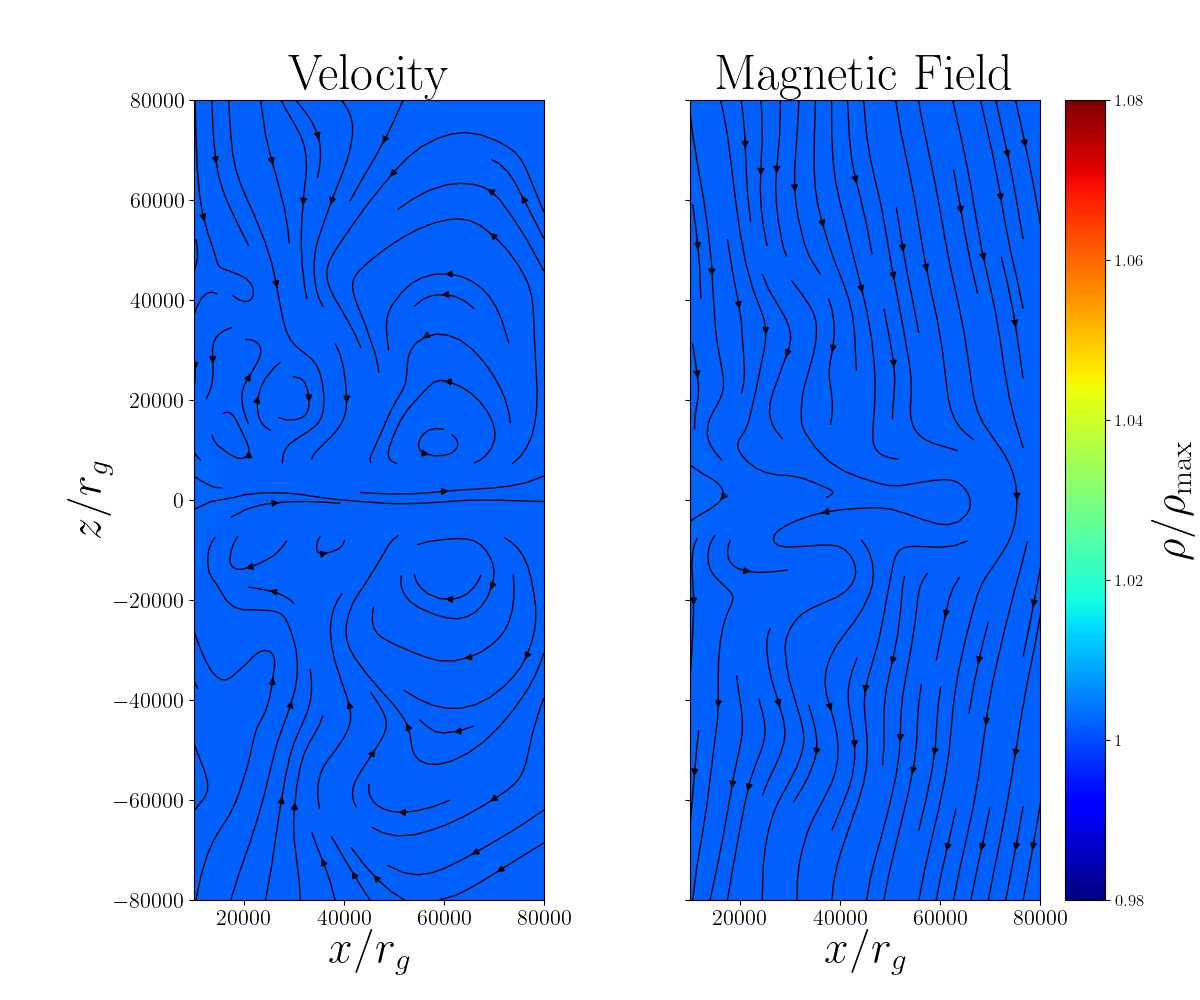}
\caption{\label{fig5} 
Velocity (left) and magnetic field (right) streamlines at $t=5\times10^7 t_g$. Colors show the normalized gas density. The MRI drives turbulence on large scales and deforms the B-field from its initial purely vertical state. The spherical average of velocity is in Fig. 1.}
\end{figure}

\section{\label{sec:level1}{GRAVITATIONAL COLLAPSE}}
The transition from distant MRI flow to thermal flow at $R = a$ defines a hierarchy of spherical accretion radii:

\begin{align}
\rho_{AMB} R_0^2 Av_K=\rho_{AMB} R_1^2c=\rho(a)a^2 c
\end{align}

\noindent Here $\rho_{AMB}$ is the ambient density undisturbed at $r > R_0$ \cite{Fowler2023}; and $R_1$ is the radius where $v$ must transition to $v = c$ to maintain flow. We see that $R_1 >> a$ requires gravitational collapse from $\rho_{AMB}$ to $\rho(a) = (R_1/a)^2\rho_{AMB} >> \rho_{AMB}$. The result is the density profile in Figure 7. 

The relevance of collapse to dark matter abundance in this paper is only to show whether a path exists to continue accretion from MRI to thermal flux at the black hole.  Why instability other than MRI (unstable sound waves etc.) is required is discussed in Reference 5 Section 4. Simulating this phase of accretion transport in KORAL would require including gravity inside the accretion zone, in addition to gravity due to M. That some velocity faster than MRI is needed is shown by $v_{MRI}t << R_0$. Effects of acceleration by pressure and magnetic field are included in the dispersion relations given in Ref. [5] (Equation (2)).\\

\textcolor{black}{\section{CREATING BLACK HOLES}}

Results for dark matter abundance in Section II only depend on MRI accretion far from M. Only gravity from mass inside a seed $M = (4\pi/c)\rho(a)a^3$ is included in KORAL, figure 1 only extends inward to $R = 100 \ R_g >> a$  so the simulation domain remains well outside the relativistic regime near M. The observed gamma ray background gives evidence that accretion extending down to $R = a$ in Figure 7 created the black holes needed to preserve dark matter to the present era. An example of how accretion created black holes is the following pressure profile for constant $\rho(a)$ matched to the Schwarzschild metric at $R > a$, giving an internal pressure $P$ at $R = 0$ given by \cite{Harris1975}:
\begin{align}
P(R = 0) = \rho(a) c_S^2 \frac{1 - \left(1 - \frac{R_S}{a}\right)^{1/2}}{3\left(1 - \frac{R_S}{a}\right)^{1/2} - 1}
\end{align}

A black hole with mass $(4\pi/3)a^3 \rho(a)$ forms when $P\rightarrow \infty$, occurring at $\ a = (9/8) R_S$. In \cite{Fowler2023}, it is noted that the average separation of accretion centers $(M/f^*\rho)^{1/3} > R_0$ for accretion fraction $f^*$, indicating that collisions of seed masses do not interfere with creating black holes.\\ 

\begin{figure}[htbp]
\includegraphics[width=0.95\columnwidth]{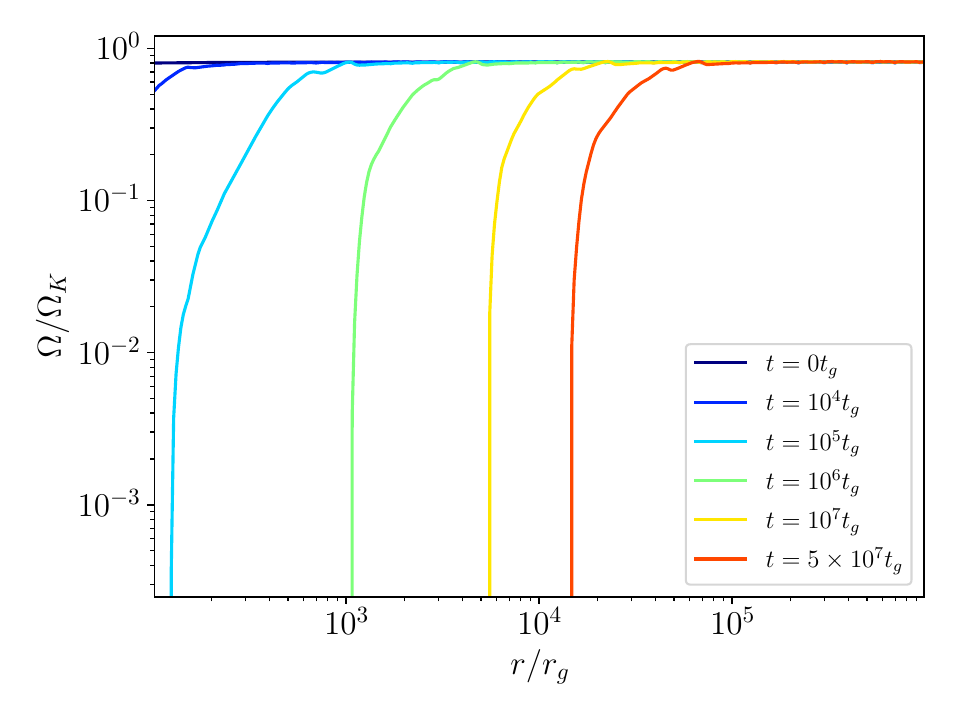}
\caption{\label{fig6} 
Showing the angular velocity $\Omega$ divided by the Keplerian angular velocity $\Omega_K$. Rotation at $R<R_0$ dies off due to angular momentum ejection. See text for description.}
\end{figure}

\begin{figure}[ht]
\includegraphics[width=\columnwidth]{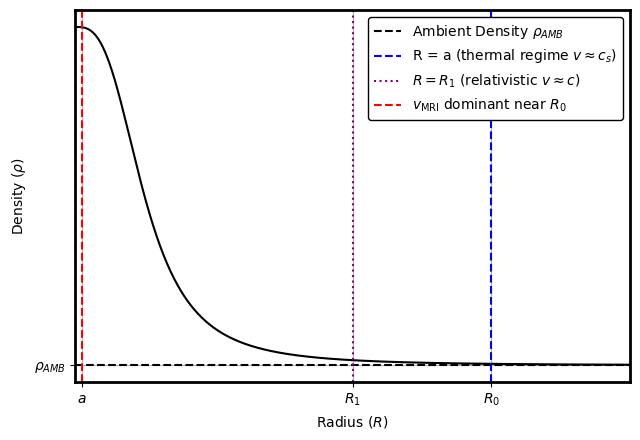}
\caption{\label{fig7} 
Showing density pile-up near the black hole, needed to sustain flow as discussed in Section V, Equation (10).
}
\end{figure}

\section{\label{sec:level1}Related Work}
\textcolor{black}{
Our work reaches several heretofore unexplored limits. Though electron-positron pair production in accretion disks has been considered, magnetic fields were not always included \cite{Sobrinho2024}. Though accretion onto primordial black holes has been considered, this has mostly been limited to Active Galactic Nuclei well after the positronium era \cite[Refs. therein]{Colgate2014, Frank2002}. Our work features accretion with rotation in a magnetic field when the matter budget of the Universe was positronium-dominated.}\\
\textcolor{black}{
\indent While others have simulated accretion as we do, our use of Figure 1 to derive accretion parameters appears to be a new application. As in other analyses, accretion requires an outward transport of angular momentum that KORAL computes but early work by Shakura and Sunyaev modeled as a phenomenological viscosity somehow associated with magnetic turbulence \cite{shakura1975black}. Similarly, early attempts to incorporate a magnetic field characterized the dynamo by a phenomenological $\alpha-\Omega$ dynamo \cite{pariev2006magnetic}. As we already noted, real progress began when Balbus and Hawley \cite{Balbus1998} introduced MRI in accretion dynamics. Even so, much attention was focused on the near neighborhood of black holes \cite{tchekhovskoy2016three}, whereas we find accretion to be dominated by MRI at a distant accretion radius $R_0$.  Also, motivated by available data, work continues on AGN’s created after the positronium era \cite{Ntormoosi2020,stone1994numerical,gao2024magnetic,Anantua2009}.}

\section{\label{sec:level1} Summary}
We have shown that assuming that dark matter was created by MRI accretion in the positronium era $0.01s < t < 14 s$ provides a unique, non-zero estimate of this source of dark matter abundance, given the accretion velocity derived by GRMHD simulations (Figure 1). Further refinement of such simulations could pin down whether and how much dark matter must arise from a “dark sector” of particle physics.

That dark matter might be black holes has a 50 year history \cite{hawking1971, Carr1975, zeldovich1967}. That black holes as dark matter must be about the size of asteroids in our solar system has been established three ways: (a) by various astronomical constraints \cite{Slatyer2024,Carr2020,Carr2021}; (b) by identifying the observed gamma ray background as Hawking radiation from black holes \cite{Carr2020, Fowler2023}; and (c) Figure 2 in this paper, establishing the range of black hole masses that would have been created by MRI-accretion of black holes from primordial positronium plasma between $0.01 - 14 \rm s$ Big Bang time. 

The main signatures of primordial black holes created in the positronium era 0.01 – 14 s after the Big Bang are the gravity of dark matter they created and their Hawking radiation. The small accretion fraction $f^* \leq 3\times 10^{-5}$ had little effect on nucleosynthesis that created the galaxies held together by this dark matter, and the pressure of their Hawking radiation was much too weak to contribute to dark energy.

The main theoretical input required for our black hole model is MRI accretion in a magnetic field. Accretion in strong magnetic fields requires some mechanism of mass flow across magnetic flux surfaces \cite{Frank2002}. A breakthrough occurred when Balbus and Hawley concluded that MRI turbulence produces magnetic jets with currents transporting mass across field lines \citep{Balbus1998}. This paper has presented GRMHD simulations using the KORAL code showing that accretion by MRI could have produced black holes contributing to dark matter. Evidence for the required primordial magnetic field is discussed in \cite{Enqvist1998}, probably created by Biermann Battery action due to primordial pressure fluctuations \cite{Kulsrud1997}. As mentioned earlier, the Peebles mechanism could have produced the rotation \citep{Peebles1969}. Carr and K\"{u}hnel \cite{Carr2020} focus on limitations on allowed masses due to constraints from astrophysical phenomena. Their constraint due to accretion refers to Bondi accretion with zero rotation. Our paper finds a second accretion window, accessed by MRI. Qualitative evidence for this from KORAL simulations is presented in Section III of this paper. While KORAL simulations exhibit MRI accretion that could produce primordial black holes, how much these black holes contribute to dark matter remains uncertain.

\textcolor{black}{As discussed in Section  II, our GRMHD simulations, together with background gamma rays \cite{Fowler2023}, indicate that $30\%$ or more of dark matter could be MRI-generated black holes, while probably leaving room for an important contribution from dark sector particles. That MRI is the likely accretion process follows from the inability of Bondi accretion to produce masses in the viable mass range in Equation (6), as shown in Section IV. Thus our work provides incentive for refinements in GRMHD simulations of MRI to extend run times as the way to better pin down how much dark sector particles ought to contribute to dark matter.}

\section{ACKNOWLEDGMENTS}
TKF acknowledges useful discussions with \textcolor{black}{Ashton Davis,} Christopher McKee and Hui Li. This work was supported by a grant from the Simons Foundation (00001470, BC, RA). Richard Anantua is supported by the Oak Ridge Associated Universities Powe Award.

\appendix
\section{Data Availability}
The data used to produce the given plots are available at \cite{Curd2025}.

\bibliography{main}

@PREAMBLE{
 "\providecommand{\noopsort}[1]{}" 
 # "\providecommand{\singleletter}[1]{#1}%" 
}

@ARTICLE{Carr2021,
       author = {{Carr}, Bernard and {Kohri}, Kazunori and {Sendouda}, Yuuiti and {Yokoyama}, Jun'ichi},
        title = "{Constraints on primordial black holes}",
      journal = {Reports on Progress in Physics},
         year = 2021,
        month = nov,
       volume = {84},
       number = {11},
          eid = {116902},
        pages = {116902},
          doi = {10.1088/1361-6633/ac1e31},
archivePrefix = {arXiv},
       eprint = {2002.12778},
 primaryClass = {astro-ph.CO},
       adsurl = {https://ui.adsabs.harvard.edu/abs/2021RPPh...84k6902C}
}

@ARTICLE{Fowler2023,
       author = {{Fowler}, T. Kenneth and {Anantua}, Richard},
        title = "{Accreting Primordial Black Holes as Dark Matter Constituents}",
      journal = {arXiv e-prints},
         year = 2023,
        month = mar,
          eid = {arXiv:2303.09341},
        pages = {arXiv:2303.09341},
          doi = {10.48550/arXiv.2303.09341},
archivePrefix = {arXiv},
       eprint = {2303.09341},
 primaryClass = {astro-ph.HE},
       adsurl = {https://ui.adsabs.harvard.edu/abs/2023arXiv230309341F}
}

@ARTICLE{carr2020,
    author = {Carr, B. and K{\"u}hnel, F.},
    year = {2020},
    journal = {Annual Review of Nuclear and Particle Science},
    volume = {70},
    pages = {355},
   doi = {10.1146/annurev-nucl-050520-125911},
     note = {see also Carr Ap. J. 1975, 201, 1; Carr et al. 2021, Reports on Progress in Physics, 84, 116902 and references therein.}
}

@ARTICLE{Carr1975,
       author = {{Carr}, B.~J.},
        title = "{The primordial black hole mass spectrum.}",
      journal = {The Astrophysical Journal},
         year = 1975,
        month = oct,
       volume = {201},
        pages = {1-19},
          doi = {10.1086/153853},
       adsurl = {https://ui.adsabs.harvard.edu/abs/1975ApJ...201....1C}
}

@article{Anantua2009,
    author = "Anantua, Richard and Easther, Richard and Giblin, John T.",
    title = "{GUT-Scale Primordial Black Holes: Consequences and Constraints}",
    eprint = "0812.0825",
    archivePrefix = "arXiv",
    primaryClass = "astro-ph",
    doi = "10.1103/PhysRevLett.103.111303",
    journal = "Phys. Rev. Lett.",
    volume = "103",
    pages = "111303",
    year = "2009"
}

@ARTICLE{Balbus1998,
       author = {{Balbus}, Steven A. and {Hawley}, John F.},
        title = "{Instability, turbulence, and enhanced transport in accretion disks}",
      journal = {Reviews of Modern Physics},
         year = 1998,
        month = jan,
       volume = {70},
       number = {1},
        pages = {1-53},
          doi = {10.1103/RevModPhys.70.1},
       adsurl = {https://ui.adsabs.harvard.edu/abs/1998RvMP...70....1B}
}

@ARTICLE{Colgate2014,
       author = {{Colgate}, Stirling A. and {Fowler}, T. Kenneth and {Li}, Hui and {Pino}, Jesse},
        title = "{Quasi-static Model of Collimated Jets and Radio Lobes. I. Accretion Disk and Jets}",
      journal = {The Astrophysical Journal},
         year = 2014,
        month = jul,
       volume = {789},
       number = {2},
          eid = {144},
        pages = {144},
          doi = {10.1088/0004-637X/789/2/144},
       adsurl = {https://ui.adsabs.harvard.edu/abs/2014ApJ...789..144C}
}

@ARTICLE{Colgate2015,
       author = {{Colgate}, Stirling A. and {Fowler}, T. Kenneth and {Li}, Hui and {Hooper}, E. and {McClenaghan}, Joseph and {Lin}, Zhihong},
        title = "{Quasi-static Model of Magnetically Collimated Jets and Radio Lobes. II. Jet Structure and Stability}",
      journal = {The Astrophysical Journal},
         year = 2015,
        month = nov,
       volume = {813},
       number = {2},
          eid = {136},
        pages = {136},
          doi = {10.1088/0004-637X/813/2/136},
       adsurl = {https://ui.adsabs.harvard.edu/abs/2015ApJ...813..136C}
}

@ARTICLE{Fowler2019,
       author = {{Fowler}, T. Kenneth and {Li}, Hui and {Anantua}, Richard},
        title = "{A Quasi-static Hyper-resistive Model of Ultra-high-energy Cosmic-ray Acceleration by Magnetically Collimated Jets Created by Active Galactic Nuclei}",
      journal = {The Astrophysical Journal},
         year = 2019,
        month = nov,
       volume = {885},
       number = {1},
          eid = {4},
        pages = {4},
          doi = {10.3847/1538-4357/ab44bc},
archivePrefix = {arXiv},
       eprint = {1903.06839},
 primaryClass = {astro-ph.HE},
       adsurl = {https://ui.adsabs.harvard.edu/abs/2019ApJ...885....4F}
}

@BOOK{Frank2002,
       author = {{Frank}, Juhan and {King}, Andrew and {Raine}, Derek J.},
        title = "{Accretion Power in Astrophysics: Third Edition}",
         year = 2002,
       adsurl = {https://ui.adsabs.harvard.edu/abs/2002apa..book.....F}
}

@ARTICLE{Cunningham2012,
       author = {{Cunningham}, Andrew J. and {McKee}, Christopher F. and {Klein}, Richard I. and {Krumholz}, Mark R. and {Teyssier}, Romain},
        title = "{Radiatively Efficient Magnetized Bondi Accretion}",
      journal = {The Astrophysical Journal},
         year = 2012,
        month = jan,
       volume = {744},
       number = {2},
          eid = {185},
        pages = {185},
          doi = {10.1088/0004-637X/744/2/185},
archivePrefix = {arXiv},
       eprint = {1201.0816},
 primaryClass = {astro-ph.SR},
       adsurl = {https://ui.adsabs.harvard.edu/abs/2012ApJ...744..185C}
}

@ARTICLE{Sadowski2013,
       author = {{Sadowski}, Aleksander and {Narayan}, Ramesh and {Tchekhovskoy}, Alexander and {Zhu}, Yucong},
        title = "{Semi-implicit scheme for treating radiation under M1 closure in general relativistic conservative fluid dynamics codes}",
      journal = {Monthly Notices of the Royal Astronomical Society},
         year = 2013,
        month = mar,
       volume = {429},
       number = {4},
        pages = {3533-3550},
          doi = {10.1093/mnras/sts632},
archivePrefix = {arXiv},
       eprint = {1212.5050},
 primaryClass = {astro-ph.HE},
       adsurl = {https://ui.adsabs.harvard.edu/abs/2013MNRAS.429.3533S}
}

@ARTICLE{Sadowski2014,
       author = {{Sadowski}, Aleksander and {Narayan}, Ramesh and {McKinney}, Jonathan C. and {Tchekhovskoy}, Alexander},
        title = "{Numerical simulations of super-critical black hole accretion flows in general relativity}",
      journal = {Monthly Notices of the Royal Astronomical Society},
         year = 2014,
        month = mar,
       volume = {439},
       number = {1},
        pages = {503-520},
          doi = {10.1093/mnras/stt2479},
archivePrefix = {arXiv},
       eprint = {1311.5900},
 primaryClass = {astro-ph.HE},
       adsurl = {https://ui.adsabs.harvard.edu/abs/2014MNRAS.439..503S}
}

@ARTICLE{Peebles1969,
       author = {{Peebles}, P.~J.~E.},
        title = "{Origin of the Angular Momentum of Galaxies}",
      journal = {The Astrophysical Journal},
         year = 1969,
        month = feb,
       volume = {155},
        pages = {393},
          doi = {10.1086/149876},
       adsurl = {https://ui.adsabs.harvard.edu/abs/1969ApJ...155..393P}
}

@ARTICLE{Enqvist1998,
       author = {{Enqvist}, Kari},
        title = "{Primordial Magnetic Fields}",
      journal = {International Journal of Modern Physics D},
         year = 1998,
        month = jan,
       volume = {7},
       number = {3},
        pages = {331-349},
          doi = {10.1142/S0218271898000243},
archivePrefix = {arXiv},
       eprint = {astro-ph/9803196},
 primaryClass = {astro-ph},
       adsurl = {https://ui.adsabs.harvard.edu/abs/1998IJMPD...7..331E}
}

@ARTICLE{Kulsrud1997,
       author = {{Kulsrud}, Russell M. and {Cen}, Renyue and {Ostriker}, Jeremiah P. and {Ryu}, Dongsu},
        title = "{The Protogalactic Origin for Cosmic Magnetic Fields}",
      journal = {The Astrophysical Journal},
         year = 1997,
        month = may,
       volume = {480},
       number = {2},
        pages = {481-491},
          doi = {10.1086/303987},
archivePrefix = {arXiv},
       eprint = {astro-ph/9607141},
 primaryClass = {astro-ph},
       adsurl = {https://ui.adsabs.harvard.edu/abs/1997ApJ...480..481K}
}

@ARTICLE{Igumenshchev2002,
       author = {{Igumenshchev}, Igor V. and {Narayan}, Ramesh},
        title = "{Three-dimensional Magnetohydrodynamic Simulations of Spherical Accretion}",
      journal = {The Astrophysical Journal},
         year = 2002,
        month = feb,
       volume = {566},
       number = {1},
        pages = {137-147},
          doi = {10.1086/338077},
archivePrefix = {arXiv},
       eprint = {astro-ph/0105365},
 primaryClass = {astro-ph},
       adsurl = {https://ui.adsabs.harvard.edu/abs/2002ApJ...566..137I}
}

@ARTICLE{McKinney2012,
       author = {{McKinney}, Jonathan C. and {Tchekhovskoy}, Alexander and {Blandford}, Roger D.},
        title = "{General relativistic magnetohydrodynamic simulations of magnetically choked accretion flows around black holes}",
      journal = {Monthly Notices of the Royal Astronomical Society},
         year = 2012,
        month = jul,
       volume = {423},
       number = {4},
        pages = {3083-3117},
          doi = {10.1111/j.1365-2966.2012.21074.x},
archivePrefix = {arXiv},
       eprint = {1201.4163},
 primaryClass = {astro-ph.HE},
       adsurl = {https://ui.adsabs.harvard.edu/abs/2012MNRAS.423.3083M}
}

@ARTICLE{Cooley2021,
       author       = "J. Cooley and others",
       title        = "Report of the Technical Group on Particle Dark Matter for Snowmass 2021",
       journal      = "arXiv",
       eprint       = "2209.07426",
       year         = "2021",
}

@ARTICLE{Slatyer2024,
   author       = "T. R. Slatyer and T. M. P. Tait",
   title        = "What If We Never Find Dark Matter?",
   journal      = "Scientific American",
   month        = "September",
   year         = "2024",
}

@article{latif2016,
  author    = {Latif, M. A. and Schleicher, D. R. G.},
  title     = {Magnetic Fields in Primordial Accretion Disks},
  journal   = {Astronomy and Astrophysics},
  year      = {2016},
  volume    = {585},
  pages     = {A151},
  doi       = {10.1051/0004-6361/201526961},
}

@article{Ntormoosi2020,
  author    = {Ntormoosi, E. and Tassis, K. and others},
  title     = {A Dynamo Amplifying the Magnetic Field in a Milky-Way-like Galaxy},
  journal   = {Astronomy and Astrophysics},
  year      = {2020},
  volume    = {641},
  pages     = {A165},
  doi       = {10.1051/0004-6361/202038585},
}

@book{bederson1999,
  editor    = {Bederson, B.},
  title     = {Reviews of Modern Physics, Special Issue 1999},
  year      = {1999},
  publisher = {American Physical Society},
  note      = {Special Issue},
}

@article{shakura1975black,
  title={Black Holes in Binary Systems. Observational Appearances},
  author={Shakura, M. I. and Sunyaev, R. A.},
  journal={A \& A},
  volume={29},
  pages={337},
  year={1975}
}

@article{pariev2006magnetic,
  title={A Magnetic Alpha-Omega Dynamo in Active Galactic Nuclei Disks: II Magnetic Field Generation, Theories and Simulations},
  author={Pariev, V. I. and Colgate, S. A. and Finn, J. M.},
  journal={arXiv:astro-ph/0611139},
  year={2006}
}

@article{tchekhovskoy2016three,
  title={Three-dimensional relativistic MHD simulations of active galactic nuclei jets; magnetic kink instability and Fanaroff-Riley dichotomy},
  author={Tchekhovskoy, A. and Bromberg, O.},
  journal={MNRAS},
  volume={461},
  pages={L46},
  year={2016}
}

@article{stone1994numerical,
  title={Numerical Simulations of Magnetic Accretion Disks},
  author={Stone, J. M. and Norman, M. L.},
  journal={ApJ},
  volume={433},
  pages={746},
  year={1994}
}

@article{gao2024magnetic,
  title={Magnetic Accretion onto and Feedback from Supermassive Black Holes in Elliptical Galaxies},
  author={Gao, M. and Stone, J. M. and Quataert, E. and Kim, C.-G.},
  journal={arXiv:2195.1171v},
  year={2024},
  month={May}
}

@article{zeldovich1967,
  author    = {Y. B. Zeldovich and I. D. Nobokov},
  title     = {Gravitational Instability of the Expanding Universe},
  journal   = {Soviet Astronomy},
  volume    = {10},
  pages     = {602},
  year      = {1967}
}

@article{hawking1971,
  author    = {S. Hawking},
  title     = {Gravitational Radiation from Colliding Black Holes},
  journal   = {Monthly Notices of the Royal Astronomical Society (MNRAS)},
  volume    = {152},
  pages     = {75},
  year      = {1971}
}

@ARTICLE{Li2013,
       author = {{Li}, Jason and {Ostriker}, Jeremiah and {Sunyaev}, Rashid},
        title = "{Rotating Accretion Flows: From Infinity to the Black Hole}",
      journal = {The Astrophysical Journal},
         year = 2013,
        month = apr,
       volume = {767},
       number = {2},
          eid = {105},
        pages = {105},
          doi = {10.1088/0004-637X/767/2/105},
archivePrefix = {arXiv},
       eprint = {1206.4059},
 primaryClass = {astro-ph.GA},
       adsurl = {https://ui.adsabs.harvard.edu/abs/2013ApJ...767..105L}
}

@ARTICLE{Hawley2011,
       author = {{Hawley}, John F. and {Guan}, Xiaoyue and {Krolik}, Julian H.},
        title = "{Assessing Quantitative Results in Accretion Simulations: From Local to Global}",
      journal = {The Astrophysical Journal},
         year = 2011,
        month = sep,
       volume = {738},
       number = {1},
          eid = {84},
        pages = {84},
          doi = {10.1088/0004-637X/738/1/84},
archivePrefix = {arXiv},
       eprint = {1103.5987},
 primaryClass = {astro-ph.HE},
       adsurl = {https://ui.adsabs.harvard.edu/abs/2011ApJ...738...84H}
}

@ARTICLE{Sadowski+2015,
       author = {{Sadowski}, Aleksander and {Narayan}, Ramesh and {Tchekhovskoy}, Alexander and {Abarca}, David and {Zhu}, Yucong and {McKinney}, Jonathan C.},
        title = "{Global simulations of axisymmetric radiative black hole accretion discs in general relativity with a mean-field magnetic dynamo}",
      journal = {Monthly Notices of the Royal Astronomical Society},
         year = 2015,
        month = feb,
       volume = {447},
       number = {1},
        pages = {49-71},
          doi = {10.1093/mnras/stu2387},
archivePrefix = {arXiv},
       eprint = {1407.4421},
 primaryClass = {astro-ph.HE},
       adsurl = {https://ui.adsabs.harvard.edu/abs/2015MNRAS.447...49S}
}

@ARTICLE{Cowling1933,
       author = {{Cowling}, T.~G.},
        title = "{The magnetic field of sunspots}",
      journal = {Monthly Notices of the Royal Astronomical Society},
         year = 1933,
        month = nov,
       volume = {94},
        pages = {39-48},
          doi = {10.1093/mnras/94.1.39},
       adsurl = {https://ui.adsabs.harvard.edu/abs/1933MNRAS..94...39C}
}

@article{Sobrinho2024,
  author = {Sobrinho, J. L. G. and Augusto, P.},
  title = {Primordial Intermediate and Supermassive Black Hole Formation During the Electron-Positron Annihilation Epoch},
  journal = {Monthly Notices of the Royal Astronomical Society},
  volume = {531},
  pages = {140},
  year = {2024},
  doi = {10.1093/mnras/stx140},
  url = {https://arxiv.org/abs/2404.07332}
}

@article{Tegmark1998,
  author = {Tegmark, M. and Rees, M. J.},
  title = {Why is the cosmic microwave background fluctuation level $10^{-5}$?},
  journal = {The Astrophysical Journal},
  volume = {499},
  pages = {526},
  year = {1998},
  doi = {10.1086/305673},
  url = {https://ui.adsabs.harvard.edu/abs/1998ApJ...499..526T}
}

@book{Harris1975,
  author = {Harris, E. G.},
  title = {Introduction to Modern Physics},
  publisher = {Wiley},
  year = {1975},
  isbn = {978-0471358802}
}

@article{Rafelski2023,
  author    = {Johann Rafelski and Jeremiah Birrell and Andrew Steinmetz and Cheng Tao Yang},
  title     = {A Short Survey of Matter-Antimatter Evolution in the Primordial Universe},
  journal   = {Universe},
  year      = {2023},
  volume    = {9},
  number    = {7},
  pages     = {309},
  doi       = {10.3390/universe9070309},
  url       = {https://arxiv.org/abs/2305.09055},
}

@article{Plank2020,
   title={Planck2018 results: VI. Cosmological parameters},
   volume={641},
   ISSN={1432-0746},
   url={http://dx.doi.org/10.1051/0004-6361/201833910},
   DOI={10.1051/0004-6361/201833910},
   journal={Astronomy and Astrophysics},
   publisher={EDP Sciences},
   author={Aghanim, N. and Akrami, Y. and Ashdown, M. and Aumont, J. and Baccigalupi, C. and Ballardini, M. and Banday, A. J. and Barreiro, R. B. and Bartolo, N. and Basak, S. and Battye, R. and Benabed, K. and Bernard, J.-P. and Bersanelli, M. and Bielewicz, P. and Bock, J. J. and Bond, J. R. and Borrill, J. and Bouchet, F. R. and Boulanger, F. and Bucher, M. and Burigana, C. and Butler, R. C. and Calabrese, E. and Cardoso, J.-F. and Carron, J. and Challinor, A. and Chiang, H. C. and Chluba, J. and Colombo, L. P. L. and Combet, C. and Contreras, D. and Crill, B. P. and Cuttaia, F. and de Bernardis, P. and de Zotti, G. and Delabrouille, J. and Delouis, J.-M. and Di Valentino, E. and Diego, J. M. and Doré, O. and Douspis, M. and Ducout, A. and Dupac, X. and Dusini, S. and Efstathiou, G. and Elsner, F. and Enßlin, T. A. and Eriksen, H. K. and Fantaye, Y. and Farhang, M. and Fergusson, J. and Fernandez-Cobos, R. and Finelli, F. and Forastieri, F. and Frailis, M. and Fraisse, A. A. and Franceschi, E. and Frolov, A. and Galeotta, S. and Galli, S. and Ganga, K. and Génova-Santos, R. T. and Gerbino, M. and Ghosh, T. and González-Nuevo, J. and Górski, K. M. and Gratton, S. and Gruppuso, A. and Gudmundsson, J. E. and Hamann, J. and Handley, W. and Hansen, F. K. and Herranz, D. and Hildebrandt, S. R. and Hivon, E. and Huang, Z. and Jaffe, A. H. and Jones, W. C. and Karakci, A. and Keihänen, E. and Keskitalo, R. and Kiiveri, K. and Kim, J. and Kisner, T. S. and Knox, L. and Krachmalnicoff, N. and Kunz, M. and Kurki-Suonio, H. and Lagache, G. and Lamarre, J.-M. and Lasenby, A. and Lattanzi, M. and Lawrence, C. R. and Le Jeune, M. and Lemos, P. and Lesgourgues, J. and Levrier, F. and Lewis, A. and Liguori, M. and Lilje, P. B. and Lilley, M. and Lindholm, V. and López-Caniego, M. and Lubin, P. M. and Ma, Y.-Z. and Macías-Pérez, J. F. and Maggio, G. and Maino, D. and Mandolesi, N. and Mangilli, A. and Marcos-Caballero, A. and Maris, M. and Martin, P. G. and Martinelli, M. and Martínez-González, E. and Matarrese, S. and Mauri, N. and McEwen, J. D. and Meinhold, P. R. and Melchiorri, A. and Mennella, A. and Migliaccio, M. and Millea, M. and Mitra, S. and Miville-Deschênes, M.-A. and Molinari, D. and Montier, L. and Morgante, G. and Moss, A. and Natoli, P. and Nørgaard-Nielsen, H. U. and Pagano, L. and Paoletti, D. and Partridge, B. and Patanchon, G. and Peiris, H. V. and Perrotta, F. and Pettorino, V. and Piacentini, F. and Polastri, L. and Polenta, G. and Puget, J.-L. and Rachen, J. P. and Reinecke, M. and Remazeilles, M. and Renzi, A. and Rocha, G. and Rosset, C. and Roudier, G. and Rubiño-Martín, J. A. and Ruiz-Granados, B. and Salvati, L. and Sandri, M. and Savelainen, M. and Scott, D. and Shellard, E. P. S. and Sirignano, C. and Sirri, G. and Spencer, L. D. and Sunyaev, R. and Suur-Uski, A.-S. and Tauber, J. A. and Tavagnacco, D. and Tenti, M. and Toffolatti, L. and Tomasi, M. and Trombetti, T. and Valenziano, L. and Valiviita, J. and Van Tent, B. and Vibert, L. and Vielva, P. and Villa, F. and Vittorio, N. and Wandelt, B. D. and Wehus, I. K. and White, M. and White, S. D. M. and Zacchei, A. and Zonca, A.},
   year={2020},
   month=sep, pages={A6} }

@dataset{Curd2025,
  author       = {Curd, Brandon},
  title        = {MRI Driven Black Holes as Dark Matter},
  year         = {2025},
  publisher    = {Zenodo},
  doi          = {10.5281/zenodo.17087686},
  url          ={https://doi.org/10.5281/zenodo.17087686}
}

@ARTICLE{Riquelme2012,
       author = {{Riquelme}, Mario A. and {Quataert}, Eliot and {Sharma}, Prateek and {Spitkovsky}, Anatoly},
        title = "{Local Two-dimensional Particle-in-cell Simulations of the Collisionless Magnetorotational Instability}",
      journal = {The Astrophysical Journal},
         year = 2012,
        month = aug,
       volume = {755},
       number = {1},
          eid = {50},
        pages = {50},
          doi = {10.1088/0004-637X/755/1/50},
archivePrefix = {arXiv},
       eprint = {1201.6407},
 primaryClass = {astro-ph.HE},
       adsurl = {https://ui.adsabs.harvard.edu/abs/2012ApJ...755...50R}
}

@ARTICLE{Sandoval2024,
       author = {{Sandoval}, Astor and {Riquelme}, Mario and {Spitkovsky}, Anatoly and {Bacchini}, Fabio},
        title = "{Particle-in-cell simulations of the magnetorotational instability in stratified shearing boxes}",
      journal = {Monthly Notices of the Royal Astronomical Society},
         year = 2024,
        month = may,
       volume = {530},
       number = {2},
        pages = {1866-1884},
          doi = {10.1093/mnras/stae959},
archivePrefix = {arXiv},
       eprint = {2308.12348},
 primaryClass = {astro-ph.HE},
       adsurl = {https://ui.adsabs.harvard.edu/abs/2024MNRAS.530.1866S}
}

@ARTICLE{Bacchini2024,
       author = {{Bacchini}, Fabio and {Zhdankin}, Vladimir and {Gorbunov}, Evgeny A. and {Werner}, Gregory R. and {Arzamasskiy}, Lev and {Begelman}, Mitchell C. and {Uzdensky}, Dmitri A.},
        title = "{Collisionless Magnetorotational Turbulence in Pair Plasmas: Steady-State Dynamics, Particle Acceleration, and Radiative Cooling}",
      journal = {Physical Review Letters},
         year = 2024,
        month = jul,
       volume = {133},
       number = {4},
          eid = {045202},
        pages = {045202},
          doi = {10.1103/PhysRevLett.133.045202},
archivePrefix = {arXiv},
       eprint = {2401.01399},
 primaryClass = {astro-ph.HE},
       adsurl = {https://ui.adsabs.harvard.edu/abs/2024PhRvL.133d5202B}
}

@ARTICLE{Rosin2012,
       author = {{Rosin}, M.~S. and {Mestel}, A.~J.},
        title = "{Quasi-global galactic magnetorotational instability with Braginskii viscosity}",
      journal = {Monthly Notices of the Royal Astronomical Society},
         year = 2012,
        month = sep,
       volume = {425},
       number = {1},
        pages = {74-86},
          doi = {10.1111/j.1365-2966.2012.21379.x},
archivePrefix = {arXiv},
       eprint = {1204.1948},
 primaryClass = {astro-ph.GA},
       adsurl = {https://ui.adsabs.harvard.edu/abs/2012MNRAS.425...74R}
}

@ARTICLE{Heinzl2022,
       author = {{Heinzl}, T.},
        title = "{QED and Lasers: A Tutorial}",
      journal = {arXiv e-prints},
         year = 2022,
        month = mar,
          eid = {arXiv:2203.01245},
        pages = {arXiv:2203.01245},
          doi = {10.48550/arXiv.2203.01245},
archivePrefix = {arXiv},
       eprint = {2203.01245},
 primaryClass = {hep-ph},
       adsurl = {https://ui.adsabs.harvard.edu/abs/2022arXiv220301245H}
}

@ARTICLE{Moeller1932,
       author = {{M{\o}ller}, Chr.},
        title = "{Zur Theorie des Durchgangs schneller Elektronen durch Materie}",
      journal = {Annalen der Physik},
         year = 1932,
        month = jan,
       volume = {406},
       number = {5},
        pages = {531-585},
          doi = {10.1002/andp.19324060506},
       adsurl = {https://ui.adsabs.harvard.edu/abs/1932AnP...406..531M}
}

\end{document}